\pdfoutput=1
\documentclass[aps,prc,onecolumn]{revtex4}
\usepackage{epsf,epsfig,amssymb,amsmath}

\newcommand{\SEL}[2]{#1}

\renewcommand{\SEL}[2]{#2}

\usepackage[dvipsnames]{xcolor}
\newcommand{\NEW}[1]{\SEL{\textcolor[named]{Blue}{#1}}{#1}}
\newcommand{\OLD}[1]{\SEL{\textcolor[named]{Red}{[#1]}}{}}

\newcommand{\be}{\begin{equation}}
\newcommand{\ee}{\end{equation}}
\newcommand{\bea}{\begin{eqnarray}}
\newcommand{\eea}{\end{eqnarray}}
\newcommand{\mybibitem}{\bibitem}

\newcommand{\gton}{\mathrel{\lower.9ex \hbox{$\stackrel{\displaystyle 
>}{\sim}$}}} 
\newcommand{\lton}{\mathrel{\lower.9ex \hbox{$\stackrel{\displaystyle 
<}{\sim}$}}}

\newcommand{\vp}{{\bf p}}

\newcommand{\myell}{{\ell}}
\newcommand{\feq}{f^{\rm eq}}

\begin{document}
%%%%%%%%%%%%%% Cover page %%%%%%%%%%%%%%%%%%%

\title{Flow harmonics from self-consistent particlization of a viscous fluid}

\author{Zack Wolff}
\author{Denes Molnar}
\affiliation{Department of Physics and Astronomy, Purdue University, West Lafayette, IN 47907}

\date{\today}

\begin{abstract}
The quantitative extraction of quark-gluon plasma (QGP) properties from heavy-ion data, such as its specific shear viscosity $\eta /s$, typically requires comparison to viscous hydrodynamic or ``hybrid'' hydrodynamics+transport simulations. In either case, one has to convert the fluid to hadrons, yet without additional theory input the conversion is ambiguous for dissipative fluids. Here, shear viscous phase-space corrections calculated using linearized transport theory are applied in Cooper-Frye freezeout to quantify the effects on anisotropic flow coefficients $v_n(p_T)$ at both RHIC and LHC energies. Expanding upon our previous flow harmonics studies \cite{MolnarWolff,Wolff:2014}, we calculate pion and proton $v_2(p_T)$, $v_4(p_T)$, and $v_6(p_T)$, but here we incorporate a hadron gas that is chemically frozen below a temperature of 175 MeV and use hypersurfaces from realistic viscous hydrodynamic simulations. For additive quark model cross sections and relative phase-space corrections with $p^{3/2}$ momentum dependence rather than the quadratic Grad form, we find at moderately high transverse momentum noticeably higher $v_4(p_T)$ and $v_6(p_T)$ for protons than for pions. In addition, the value of $\eta /s$ deduced from elliptic flow data differs by nearly 50\% from the value extracted using the naive ``democratic Grad'' form of freeze-out distributions. To facilitate the use of the self-consistent viscous corrections calculated here in hydrodynamic and hybrid calculations, we also present convenient parameterizations of the corrections for the various hadron species (cf. Table \ref{Table:Cfits-Tch175}).

\end{abstract}

\maketitle

%%%%%%%%%%%%%
\section{Introduction}
\label{Sc:Intro}

The most widely used framework for describing the early stages of a heavy-ion collision is relativistic hydrodynamics \NEW{\cite{Huovinen:2006,Teaney:2009,Gale:2013}}\OLD{[3]}. The calculation of heavy-ion observables from a hydrodynamic simulation requires the conversion of an expanding fluid into a description in terms of hadrons\NEW{, often}\OLD{. This is} referred to as particlization \cite{HuovinenPetersen}. The hadrons are then either evolved further in a transport model or assumed to free-stream to the detectors. The conversion is usually done using the Cooper-Frye \cite{CooperFrye} prescription that gives the distribution of particles emitted from the fluid across a hypersurface in spacetime (typically chosen as a constant-temperature \NEW{or energy density} hypersurface). This requires knowledge of the local distribution functions of each hadron species in momentum space. If the fluid is in local thermal equilibrium, i.e., an ideal fluid, then the distributions are uniquely determined by the hydrodynamic variables \NEW{\cite{MolnarWolff,dyngrad}}. For viscous fluids, however, an infinite number of particle distributions will match the hydrodynamic fields \NEW{(see discussion in Sec. II)}; therefore, additional theory input is required.

The ambiguity in the viscous particle distributions is often ignored, and in practice, relative corrections to thermal distributions are assumed to be quadratic in momentum (Grad ansatz). Moreover, the distributions are commonly taken to be independent of the hadron scattering rates that are responsible for keeping the gas near equilibrium, which we refer to as the ``democratic Grad'' ansatz \cite{dyngrad}. In Refs. \cite{MolnarWolff,Wolff:2014}, the ambiguity was resolved self-consistently by calculating the distributions from the linearized Boltzmann equation, and shear viscous corrections proportional to $p^{3/2}$ power of momentum were found to be favored over the quadratic Grad dependence. Here we expand upon those works by implementing an equation of state for a hadron gas chemically frozen below 175 MeV, as well as hypersurfaces obtained from real viscous hydrodynamic evolution. In addition, we estimate the uncertainty in the specific shear viscosity of the quark-gluon plasma (QGP) deduced from elliptic flow data, and study how the fluid-to-particle conversion affects higher anisotropic flow coefficients $v_4(p_T)$ and $v_6(p_T)$ at both RHIC and LHC energies.

\NEW{The paper is structered as follows: In Section II we summarize the self-consistent approach to calculating shear viscous particle distributions and collect the major results from Ref. \cite{MolnarWolff}. In Section III A and B particle distributions are calculated for a chemically frozen effective hadron gas. These distributions are then used in Cooper-Frye freezeout in Section III C to calculate differential elliptic flow. The uncertainty in the specific shear viscosity in the hybrid approach is quantified in Section III D and in Section III E we compare the value obtained for the specific shear viscosity to results from other theoretical frameworks. In Section III F we conclude with results for higher anisotropic flow coefficients $v_4(p_T)$ and $v_6(p_T)$.}

%%%%%%%%%%
\section{Cooper-Frye procedure and viscous phase-space corrections}
\label{Sc:framework}

After hydrodynamic evolution, the distribution of \NEW{particles emitted by the fluid is typically calculated using the Cooper-Frye \cite{CooperFrye} prescription. The number $N_i$ of particles of species $i$ with 4-momentum $p^\mu$ emitted from a fluid surface element $d\sigma_\mu$ of a 3D hypersurface embedded in 4D spacetime at point $x$ is}
\OLD{particle species $i$ emitted from a fluid surface element $d\sigma^\mu$ of a 3D hypersurface embedded in 4D spacetime is obtained using the Cooper-Frye prescription}
\be
E_i\frac{dN_i(x,\vp)}{d^3p} = p^\mu d\sigma_\mu(x)  f_i(x,\vp) \ ,
\label{CooperFrye}
\ee
\NEW{where $E_i = \sqrt{p^2 + m_i^2}$ is the on-shell energy of a particle of mass $m_i$.} The procedure requires not only knowledge of the constant-temperature hypersurface given by the hydrodynamic simulation, but also the full phase-space distribution functions of emitted particle species in momentum space, $f_i(x,\vp)$. The latter can be separated into a local equilibrium part and a viscous correction, $f_i \equiv f_i^{eq} + \delta f_i$.

For ideal fluids in local equilibrium, it is straightforward to obtain the equilibrium distributions of outgoing particles from the energy-momentum tensor. In a fluid with shear viscosity, however, there are in general shear corrections $\pi^{\mu\nu}(x)$ to the ideal energy-momentum tensor. To find the viscous corrections to thermal distribution functions, one must invert
\be
\pi^{\mu\nu}(x) = \sum\limits_i \int\limits \frac{d^3 p}{E} p^\mu p^\nu \delta f_i(x,\vp) \ .
\label{constraint}
\ee
The problem is that infinitely many viscous correction functions $\delta f_i(x,\vp)$ satisfiy the constraint (\ref{constraint}), even if there is only a single particle species present. \NEW{The democratic Grad ansatz, which is quadratic in momentum, is but one ad-hoc choice from among these. One could pick, for example, arbitrary power-law momentum dependence instead \footnote{\NEW{One such infinite class which satisfies constraint (\ref{constraint}) is the generalization of the quadratic $p^2$ Grad form to a general power law $p^\alpha$:
$$
\delta f_\alpha \equiv c_\alpha \ \left(\frac{p \! \cdot \! u}{T}\right)^{\alpha -2} \frac{\pi^{\mu\nu} p_\mu p_\nu}{2(e+P)T^2} f^{eq} \ , 
$$ 
where the normalization constant
$$ 
c_\alpha = \frac{15 z^3 K_3(z)}{I_\alpha(z)} \ , \qquad {\rm with} \qquad  I_\alpha(z) \equiv \int\limits_z^\infty dx \ x^{\alpha - 2} (x^2-z^2)^{5/2} e^{-x}
$$
is fixed by the requirement that $\delta f_\alpha$ reproduces the given local shear stress $\pi^{\mu\nu}$. Setting $\alpha = 2$ reproduces the standard Grad coefficient with $c_\alpha = 1$.}}.} 

The ambiguity was resolved in Ref. \cite{MolnarWolff} by calculating the distribution functions using the linearized Boltzmann equation. In this way, the collision rates between particles that keep the hadron gas near equilibrium are taken into account, and no ad-hoc assumptions about the momentum dependence of the corrections are necessary \footnote{While no assumption about the momentum dependence is needed a priori, for simplicity, the relative corrections are \NEW{here} taken to be proportional to a single power of momentum, though not necessarily quadratic.}. 
Here we recapitulate key ingredients of that approach (the reader is directed to Ref. \cite{MolnarWolff} and references therein for more detail). First, the viscous corrections $\delta f_i$ can be reduced to a dimensionless function of momentum, $\chi_i(|\tilde\vp|)$, defined by
\be
\delta f_i / f_i^{eq} \equiv \phi_i(x,\vp) \equiv \chi_i(|\tilde\vp|) P^{\mu\nu} X_{\mu\nu}
\quad \quad {\rm with} \quad
\left.\frac{1}{T}\Delta^{\mu\nu} p_\nu\right|_{LR} \equiv (0, \tilde\vp) \ ,
\label{chi_def} 
\ee
where $\Delta^{\mu\nu} \equiv g^{\mu\nu} - u^\mu u^\nu$ is a spatial projector perpendicular to the flow velocity $u^\mu$ such that $\tilde\vp$ is the three-momentum in the local fluid rest frame (LR) normalized by temperature. The tensors 
\be
P^{\mu\nu} \equiv \frac{1}{T^2}\left[
        \Delta^\mu_\alpha \Delta^\nu_\beta p^\alpha p^\beta - 
          \frac{1}{3}\Delta^{\mu\nu} (\Delta_{\alpha\beta} p^\alpha p^\beta)
        \right]
\quad \ {\rm and} \quad \
X^{\mu\nu} \equiv \frac{1}{T} (\nabla^\mu u^\nu + \nabla^\nu u^\mu 
                 - \frac{2}{3} \Delta^{\mu\nu} \partial_\alpha u^\alpha) , 
\quad (\nabla^\mu \equiv \Delta^{\mu\nu} \partial_\nu )
\label{def_P_X}
\ee
are dimensionless, symmetric, traceless, and purely spatial in the LR frame. The self-consistent phase-space corrections $\{\chi_i\}$ are then given by a linear integral equation, which can be solved via extremizing the functional
\bea
Q[\chi] &\equiv& \frac{1}{2T^2} \sum\limits_i \int\limits_1 
                            P_1 \cdot P_1 \feq_{1i} \chi_{1i}
\nonumber\\
         && +\ \frac{1}{2T^4} \sum\limits_{ijk\myell}
                        \int\limits_1\!\!\!\!\int\limits_2\!\!\!\!
                        \int\limits_3\!\!\!\!\int\limits_4
           \feq_{1i} \feq_{2j} \, \bar W_{12\to 34}^{ij\to k\myell}\,
           \delta^4(12-34)\,
          (  \chi_{3k} P_3 \cdot P_1 
                          + \chi_{4\myell} P_4 \cdot P_1
                          - \chi_{1i} P_1 \cdot P_1
                          - \chi_{2j} P_2 \cdot P_1) \chi_{1i} \ .
\label{Qdef}
\eea
Here,
\be
\int\limits_a \equiv \int d^3p_a / (2 E_a) \ , \quad  
P_a \cdot P_b \equiv P_a^{\mu\nu}P_{b,\mu\nu} \ , \quad
\chi_{ai} \equiv \chi_i(|\tilde \vp_a|) \quad , \quad
\delta^4(ab - cd) \equiv \delta^4(p_a + p_b - p_c - p_d) \ ,
\ee
where $a$, $b$, $c$, and $d$ label particle momenta in microscopic $2 \to 2$ scatterings. Once the transition probabilities $\bar W_{12\to 34}^{ij\to k\myell}$ are specified, one can evaluate the functional in Eq. (\ref{Qdef}) (if necessary, numerically), and obtain a variational solution for the viscous corrections $\chi_i(|\tilde \vp|)$ for all particle species in the system.

%%%%%%%%%%
\section{Hadron gas and anisotropic flow}
\label{Sc:HG_flow}

The self-consistent method described in Section \ref{Sc:framework} will now be applied to a gas of hadrons near thermal equilibrium to calculate the dissipative corrections at particlization based on the microscopic dynamics of each species. Unlike in Ref. \cite{MolnarWolff}, where all hadrons in the gas were taken to be in chemical equilibrium, shear corrections are now computed for a hadron gas that is chemically frozen for temperatures $T < T_{ch} = 175$~MeV.

\subsection{Chemical freezeout in an effective hadron gas}
\label{Sc:HG_coeffs}

The dynamics of a realistic hadron gas are complicated, as each species has a unique energy-dependent (and possibly angle-dependent) cross section with each other species in the system. Here a simple model of interactions is considered with constant meson-meson, meson-baryon, and baryon-baryon cross sections in the proportions $\sigma_{MM} : \sigma_{MB} : \sigma_{BB} = 4:6:9$, motivated by the additive quark model (AQM) \cite{UrQMD,AQM}. The overall magnitudes of the cross sections are set by the shear viscosity of the system. Only elastic $ij \to ij$ scattering (allowing $i=j$) is considered with isotropic, energy-independent cross sections. In this way, a simple model can be investigated that still includes hadronic species dependence. It has been shown previously \cite{MolnarWolff,DenicolQM2012} that if one postulates the same constant cross sections for all particle species, then the results for heavy-ion observables are very similar to those of the democratic Grad ansatz typically employed. As in Ref. \cite{MolnarWolff}, to simplify the calculation, we combine members of the same isospin multiplet, as well as their antiparticles, into a single effective species with an appropriately scaled degeneracy factor. Hadrons up to mass 1.672 GeV, i.e., the $\Omega(1672)$, are included in this way, yielding a mixture of 49 effective species. 

The cross sections for inelastic, particle-number-changing processes are known to be smaller than those of elastic, resonance-forming processes at lower temperatures \cite{Goity:1989,Gerber:1990}. Final hadron abundance ratios also seem to favor a chemical freeze-out temperature of about $T_{ch} \approx 160-175$ MeV \cite{Braun-Munz:2004,STAR:2005}, while the slopes of spectra prefer a lower kinetic freeze-out temperature \cite{STAR:2004}. Therefore, unlike in Ref. \cite{MolnarWolff}, we now allow for separate kinetic $T_{FO}$ and chemical $T_{ch}$ freeze-out temperatures with $T_{ch} \ge T_{FO}$. Following the approach of Ref. \cite{Bebie:1992}, temperature-dependent chemical potentials are introduced for each species such that relative abundances of species are locked in for $T < T_{ch}$ to their values at $T_{ch}$:
\be
\frac{n_i(T,\mu_i)}{n_j(T,\mu_j)} = \frac{n_i(T_{ch},0)}{n_j(T_{ch},0)} \ .
\label{nchemeq}
\ee
The $49-1=48$ independent ratio equations summarized in Eq. (\ref{nchemeq}) allow one to write all chemical potentials of the effective hadron gas in terms of one of the chemical potentials, for example, $\mu_\pi$. The last chemical potential is then fixed by requiring that the ratio of particle density to entropy density remains unchanged along flow streamlines as in Ref. \cite{Huovinen:2008}:
\be
\frac{n_\pi(T,\mu_\pi)}{s(T,\{\mu_i\})} = \frac{n_\pi(T_{ch},0)}{s(T_{ch},\{\mu_i=0\})} \qquad \qquad \qquad (T<T_{ch}) \ .
\label{nschemeq}
\ee

With chemical potentials calculated from Eqs. (\ref{nchemeq}) and (\ref{nschemeq}), we solve the variational problem numerically using the same adaptive integration routines from the GNU Standard Library (GSL) \cite{GSL} as in Ref. \cite{MolnarWolff}. For simplicity, as in Ref. \cite{MolnarWolff}, power-law dependence is considered here with relative viscous corrections that are either quadratic in momentum (Grad case) or proportional to $p^{3/2}$. The exponent 3/2 is motivated by the variational solutions for a gas of hadrons in \cite{MolnarWolff}, as well as analytical results for massless species \cite{thesis}. It is useful to define viscous correction coefficients $c_i$ by factoring out the momentum dependence as
\be
\chi_i^{Grad} = c^{Grad}_i |\tilde\vp|^0 \chi^{dem} \ ,
\qquad\qquad \chi_i^{(3/2)}(|\tilde\vp|) = c^{(3/2)}_i |\tilde\vp|^{-1/2} \chi^{dem} \ ,
\label{df_corrections}
\ee
where 
\be
\chi^{dem} = \frac{\eta T}{2(e+P)} = \frac{1}{2} \frac{\eta}{s + \sum\limits_{c} \mu_c n_c / T} 
\label{chi_dem}
\ee
corresponds to the species-independent democratic Grad corrections expressed in terms of thermal values of pressure $P$, energy density $e$, and charge densities $n_c$. For example, $c^{Grad}_i$ quantifies how far species $i$ is from the democratic form. Note that in the limit of vanishing chemical potentials, $\mu_c \rightarrow 0$, the viscous corrections are proportional to the shear viscosity to entropy density ratio $\eta / s$.

To facilitate inclusion of the self-consistent viscous corrections calculated here in hydrodynamic and hybrid calculations, in Table \ref{Table:Cfits-Tch175} we list convenient parameterizations of the corrections for the various hadron species for  the two power-law scenarios at conversion temperatures $T_{conv} = 100$, 120, 140, and $160$~MeV. \NEW{In order to match the hydrodynamic equation of state used in the elliptic flow study in Sec. III C, we set $T_{ch} = 175$~MeV.}
Despite variations in the degeneracy factors between hadronic species, the viscous correction coefficients $c_i$ depend on hadron mass rather smoothly, and can be fit well with
\be
c(x) = \delta + \alpha \left[1 + \displaystyle{\left(\frac{x}{\gamma}\right)^\beta}\right]^{-1} \ , 
\qquad x \equiv \frac{m}{1\ {\rm GeV}} \ ,
\label{Cfit_fns}
\ee 
where $x$ is the hadron (pole) mass $m$ in GeV. The functional form (\ref{Cfit_fns}) was chosen empirically for accuracy (the relative accuracy of the fits is better than $10^{-3}$), i.e., it does not reflect any physics motivation. To apply the dynamical correction for species $i$, take the appropriate coefficient $c_i$ from the table and multiply democratic viscous corrections by the expression in (\ref{df_corrections}) that corresponds to the assumed momentum dependence. 

%%%%%%%%%
%fit functions - Grad and p^3/2
\begin{table}
\begin{center}
\begin{tabular}{ccccc}
\hline\hline \\[-0.3cm]
\multicolumn{5}{l}{fits for AQM cross sections with $T_{ch}=175$~MeV} \\[0.2cm]
\multicolumn{5}{l}{
  using $c(x) = \delta
          + \alpha \left[1 + \displaystyle{\left(\frac{x}{\gamma}\right)^\beta                             }\right]^{-1}$ } \\[0.5cm]
\hline\hline \\[-0.2cm]
\multicolumn{5}{c}{$\delta f/f_{eq} \propto p^2$ (Grad), mesons} \\[0.1cm]
$T$~[MeV] & $\alpha$ & $\beta$ & $\gamma$ & $\delta$ \cr
\hline
       100 & 1.001 & 1.237 & 0.824 & 0.555  \cr
       120 & 0.894 & 1.302 & 0.931 & 0.572  \cr
       140 & 0.815 & 1.359 & 1.026 & 0.587  \cr
       160 & 0.752 & 1.407 & 1.112 & 0.600  \cr
\hline \\[-0.2cm]
\multicolumn{5}{c}{$\delta f/f_{eq} \propto p^2$ (Grad), baryons} \\[0.1cm]
$T$~[MeV] & $\alpha$ & $\beta$ & $\gamma$ & $\delta$ \cr
\hline
       100 & 0.955 & 1.014 & 0.784 & 0.317  \cr
       120 & 0.867 & 1.052 & 0.925 & 0.323  \cr
       140 & 0.798 & 1.089 & 1.061 & 0.330  \cr
       160 & 0.742 & 1.124 & 1.190 & 0.337  \cr
\hline \\[-0.2cm]
\multicolumn{5}{c}{$\delta f/f_{eq} \propto p^{3/2}$, mesons} \\[0.1cm]
$T$~[MeV] & $\alpha$ & $\beta$ & $\gamma$ & $\delta$ \cr
\hline
       100 & 1.361 & 1.261 & 0.783 & 2.331  \cr
       120 & 1.215 & 1.308 & 0.879 & 2.265  \cr
       140 & 1.104 & 1.350 & 0.967 & 2.217  \cr
       160 & 1.018 & 1.388 & 1.048 & 2.180  \cr
\hline \\[-0.2cm]
\multicolumn{5}{c}{$\delta f/f_{eq} \propto p^{3/2}$, baryons} \\[0.1cm]
$T$~[MeV] & $\alpha$ & $\beta$ & $\gamma$ & $\delta$ \cr
\hline
       100 & 1.127 & 1.229 & 0.897 & 1.611  \cr
       120 & 1.055 & 1.225 & 0.987 & 1.552  \cr
       140 & 0.994 & 1.227 & 1.083 & 1.507  \cr
       160 & 0.942 & 1.233 & 1.182 & 1.473  \cr
\end{tabular}
\end{center}
\caption{Parametrization of the species-dependent shear viscous corrections in a 49-species hadron gas that is chemically frozen below $T_{ch} = 175$~MeV, with additive quark model cross sections (see text), and assuming either quadratic (top two tables) or $p^{3/2}$ (bottom two tables) momentum dependence for the relative correction $\delta f / f^{eq}$.}
\label{Table:Cfits-Tch175}
\end{table}

\subsection{Coefficients $c_i$ in the dynamic Grad approximation}

It is instructive to look at the overall effect of chemical freezeout on the viscous corrections. For corrections that are quadratic in momentum, the viscous coefficients of pions and heavier resonances are higher in the chemically frozen case. For example, at a conversion temperature of $100$~MeV the relative correction for pions and protons are $c_\pi = 1.46$, $c_N = 0.75$ with chemical freezeout at $175$~MeV, while $c_\pi = 1.08$, $c_N = 0.56$ in full chemical equilibrium at $T_{conv}$. Early chemical freezeout brings the pions further above the democratic baseline of $c_\pi = 1$, and the heavier species closer to the democratic baseline ($c_i = 1$). The reason why pions go further out of equilibrium is that, due to the chemical potentials, their relative abundance decreases for temperatures below $T_{ch}$.

Even after incorporating chemical freezeout, the ratio of pion to proton viscous correction coefficients stays $c^{Grad}_\pi / c^{Grad}_N \approx 2$ for $100 < T_{conv} < 160$~MeV. Therefore, just like in chemical equilibrium \cite{MolnarWolff}, protons are still about twice as equilibrated thermally as pions, reflecting the larger overall scattering cross sections for baryons relative to mesons in the additive quark model. The pion-proton difference, therefore, will also manifest in identified particle observables when self-consistent, species-dependent distribution functions are included in Cooper-Frye freezeout.

%%%%%%%%
\subsection{Elliptic flow for $\delta f \propto p^{3/2}$}
\label{Sc:v2pto3o2}

From here on, we focus on relative shear corrections with $p^{3/2}$ momentum dependence (i.e., $\phi_i \equiv \delta f_i / f^{eq}_i \propto p^{3/2}$). 
To quantify the effect of self-consistent freezeout in heavy-ion collisions at RHIC and LHC energies, we first calculate differential elliptic flow $v_2(p_T,y)$, defined as the second Fourier coefficient of the azimuthal momentum distribution at fixed transverse momentum $p_T$ and rapidity $y$:
\be
E \frac{d^3N}{d^3p} = \frac{1}{2\pi}\frac{d^2N}{p_Tdp_Tdy}\left(1+2\sum\limits_{n=1}^\infty v_n(p_T,y)\cos[n(\phi-\Psi_{n,RP})]\right) \ .
\label{vn}
\ee
Here $\phi$ is the azimuthal angle around the beam axis and $\Psi_{n,RP}$ is the reaction-plane angle for the $n$-th harmonic.
Specifically we calculate $v_2(p_T)$ for $Au + Au$ collisions at top RHIC energy $\sqrt{s_{NN}}=200$~GeV, and for $Pb + Pb$ collisions at the LHC at $\sqrt{s_{NN}}=2.76$~TeV, for 20-30\% centrality in both cases. Hypersurface data from boost-invariant 2+1D viscous hydrodynamic simulations from smooth $A+A$ initial conditions %
\footnote{The initial conditions are those used in Ref. \cite{PaatelainenNiemi:2014} with saturation coefficient $K_{sat} = 0.69$, hardness parameter $\beta = 0.9$, and Bjorken scaling of the energy density before thermalization (``BJ'' scenario).}, 
using the realistic QCD equation of state parameterization s95p-PCE-v1 \cite{HuovinenPetreczky:2010} with chemical freezeout at $T_{ch}=175$~MeV, were obtained from H. Niemi (the same hypersurface data were also used in Ref. \cite{PaatelainenNiemi:2014}). In the hydro calculations, the specific shear viscosity of the system was constant $\eta / s = 0.12$ in the deconfined phase, while it decreased linearly with temperature in the hadron gas phase (cf. Fig. 2 of Ref. \cite{PaatelainenNiemi:2014}). For both RHIC and LHC collisions, the kinetic freezeout 
temperature was chosen to be $T_{FO} = 100$~MeV. For further details on the simulation parameters, see Ref. \cite{PaatelainenNiemi:2014}. Details of the numerical algorithm used in the simulations can be found in Refs. \cite{Niemi:2011,EMolnarNiemi:2010}. The viscous Cooper-Frye integrals were evaluated numerically %
\footnote{In hydrodynamic applications one uses the local shear stress tensor directly, i.e., in the viscous correction Eq. (\ref{chi_def}) substitutes $X_{\mu\nu} \to \pi_{\mu\nu} / \eta T$.}
the same way as in Ref. \cite{MolnarWolff}. 
After the fluid-to-particle conversion, unstable resonances in the system were decayed to pions, protons, and kaons using the {\sf RESO} algorithm in the {\sf AZHYDRO} package \cite{AZHYDRO,AZHYDROrefs}.

\begin{figure}[h!]
\includegraphics[width=0.49\linewidth]{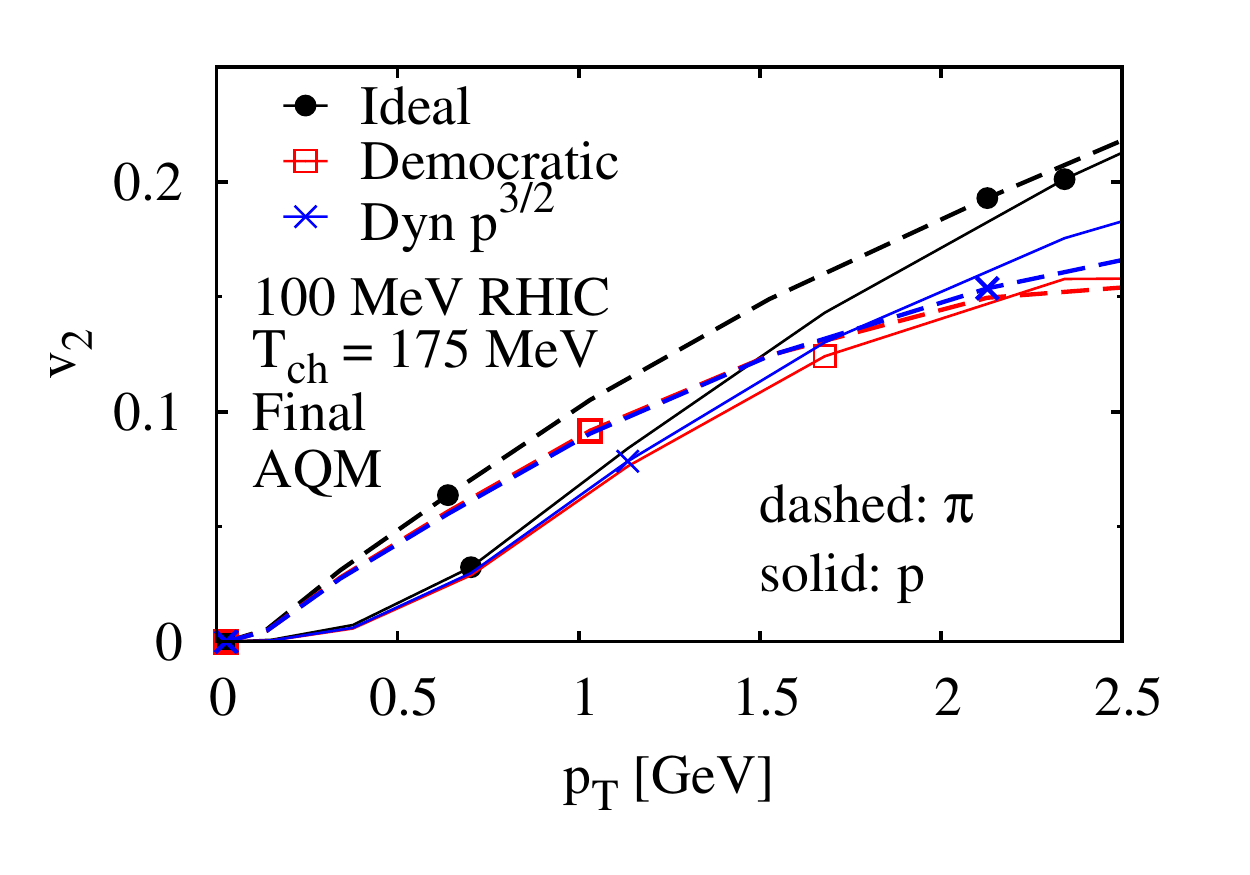}
\caption{Differential elliptic flow $v_2(p_T)$ of pions and protons in $Au + Au$ at $\sqrt{s_{NN}}=200$~GeV at 20-30\% centrality using 2+1D boost invariant viscous hydrodynamic solutions \cite{PaatelainenNiemi:2014} and fluid-to-particle conversion at $T_{conv} = 100$~MeV. Feed-down from decays of unstable resonances has been included. Dashed lines are for pions, while solid curves are for protons. The standard democratic Grad approach (open boxes) is compared to self-consistent shear corrections with momentum dependence $\delta f_i \propto p^{3/2}$ (crosses) computed for a 49-species effective hadron gas from linearized kinetic theory (see text). Results with uncorrected, local equilibrium phase-space distributions ($\delta f = 0$) are also shown (filled circles) for reference.}
\label{Fig:RHIC-v2-3o2}
\end{figure}

Figure \ref{Fig:RHIC-v2-3o2} compares pion and proton differential elliptic flow for  $Au + Au$ at RHIC with fluid-to-particle conversion at $T_{conv} = 100$~MeV using the commonly employed democratic Grad ansatz (open boxes) and the self-consistent approach (crosses). For reference, results from freezeout without any viscous corrections ($\delta f = 0$) are also shown (filled circles).  
 In the democratic Grad scenario, dissipation reduces elliptic flow by around 35\% for both species at higher $p_T$ compared to ideal, nonviscous, freezeout. In contrast, with self-consistent, species-dependent freezeout, protons are closer to equilibrium and, therefore, proton elliptic flow is much less suppressed at high $p_T$. On top of this effect, there is also an increase in $v_2$ for both species at larger $p_T$ due to the weaker $\delta f \propto p^{3/2}$ momentum dependence compared to the quadratic one assumed in the democratic Grad case.
The mass ordering of elliptic flow, $v_2^p < v_2^\pi$, is also present at low $p_T$ in all freeze-out scenarios, characteristic of a common hydrodynamic velocity for all species. For viscous freezeout the mass ordering reverses at higher $p_T$, so the pion and proton curves necessarily cross. With self-consistent viscous corrections, the crossing occurs at significantly lower $p_T \approx 1.6$~GeV because protons are closer to equilibrium (hence, their $v_2$ is larger) than in the democratic Grad case. These features combine to give a pion-proton elliptic flow splitting of roughly 10\% at $p_T \approx 2.5$~GeV with self-consistent particlization, compared to a neglible splitting if one uses the democratic Grad ansatz. 

At the low conversion temperatures $T \approx 100$~MeV used here, realistic viscous hydrodynamic evolution in fact gives rise to much larger viscous corrections at high $p_T$ than the Navier-Stokes shear stress estimate
\be
\pi^{\mu\nu} = \eta \left(\nabla^\mu u^\nu + \nabla^\nu u^\mu - \frac{2}{3}\Delta^{\mu\nu} \partial_\alpha u^\alpha \right)
\label{pi_est}
\ee
that was used in Refs. \cite{MolnarWolff,Wolff:2014}. By late times, flow gradients get smoothed out so effectively that the calculation in Ref. \cite{MolnarWolff} would give negligible shear corrections to both pion and proton $v_2(p_T)$, even with the three times larger $\eta / s \approx 0.3$ at $T_{conv} = 100$~MeV in the simulations used here. The influence of early chemical freezeout on identified elliptic flow is, however, smaller. It leads to a roughly 5\% relative increase for pion $v_2$, while it leaves proton flow practically unaffected.   

\begin{figure}[h!]
\includegraphics[width=0.49\linewidth]{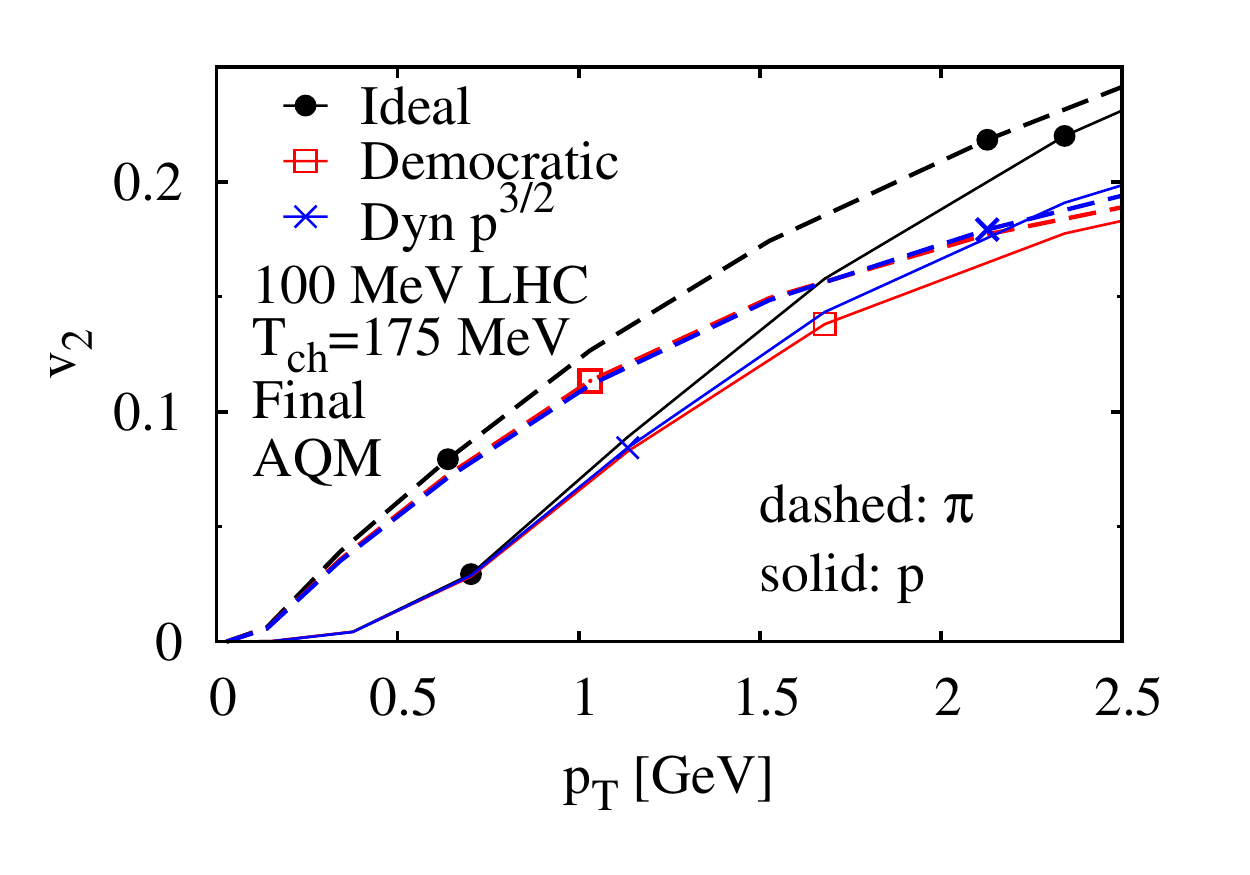}
\includegraphics[width=0.49\linewidth]{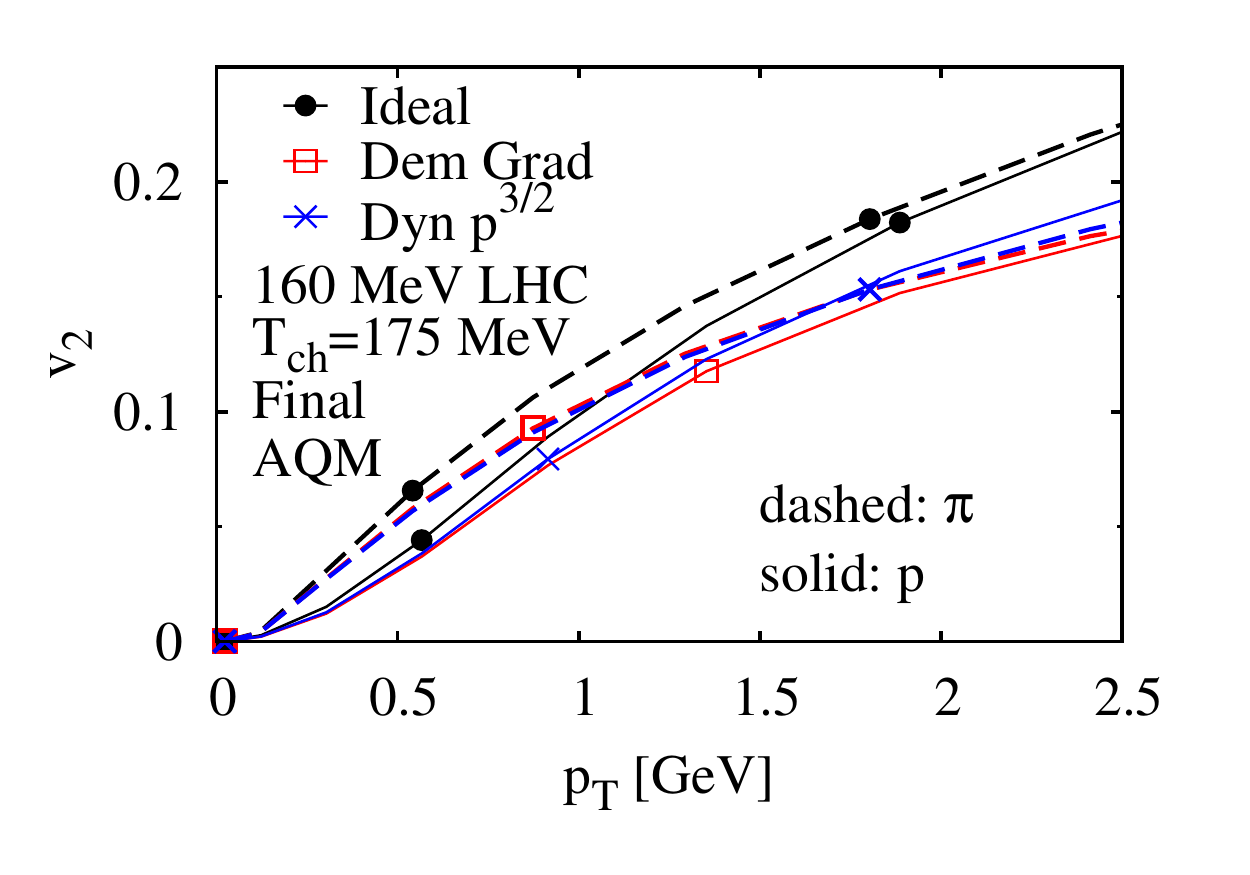}
\caption{Differential elliptic flow $v_2(p_T)$ of pions and protons in $Pb + Pb$ at $\sqrt{s_{NN}}=2.76$~TeV at the LHC at 20-30\% centrality using 2+1D boost invariant viscous hydrodynamic solutions\cite{PaatelainenNiemi:2014} and fluid-to-particle conversion at $T_{conv} = 100$~MeV (left panel) or 160 MeV (right panel). Feed-down from decays of unstable resonances has been included. Dashed lines are for pions, while solid curves are for protons. The standard democratic Grad approach (open boxes) is compared to self-consistent shear corrections with momentum dependence $\delta f_i \propto p^{3/2}$ (crosses) computed for a 49-species effective hadron gas from linearized kinetic theory (see text). Results with uncorrected, local equilibrium phase-space distributions ($\delta f = 0$) are also shown (filled circles) for reference.}
\label{Fig:LHC-v2-3o2}
\end{figure}

To investigate the influence the collision energy of the system has on the elliptic flow, the left panel of Fig. \ref{Fig:LHC-v2-3o2} shows the analogous calculation for midcentral $Pb + Pb$ collisions at the LHC at $\sqrt{s_{NN}} = 2.76$~TeV. The initial temperature of the fireball is now higher, so the system evolves longer by the time it reaches the $T_{conv} = 100$~MeV hypersurface. This leads to an increase in both proton and pion elliptic flow, as well as a more pronounced mass splitting at low $p_T$. As a result, the viscous curves cross at noticeably higher $p_T \approx 2.2$~GeV with self-consistent conversion (crosses), while at $p_T > 2.5$~GeV for democratic freezeout (open boxes).

To test the sensitivity of the elliptic flow to the assumed conversion temperature, the same observables in $Pb + Pb$ collisions at the LHC were also calculated using hypersurfaces at higher $T_{conv} = $ 120, 140, and 160 MeV. Here we only compare results for $T_{conv} = 100$ (Fig. \ref{Fig:LHC-v2-3o2} left panel) and $160$~MeV (Fig. \ref{Fig:LHC-v2-3o2} right panel) because the intermediate temperatures qualitatively interpolate between those two extremes. At $T_{conv} = 160$~MeV, there is only a slight reduction in the differential elliptic flow for pions and protons but the mass effect at low $p_T$ is greatly reduced because it is driven by the difference in $m/T$ between the species. While the pion-proton flow crossing shifts from  $p_T \approx 2.2$~GeV to $\approx 1.6$~GeV for self-consistent viscous corrections (crosses), the difference between pion and proton elliptic flow at high $p_T \approx 2.5$~GeV is insensitive to the conversion temperature.

%%%
\subsection{Uncertainty in shear viscosity extraction}
\label{Sc:Extract}

The results in the previous section highlight the sensitivity of identified particle flow harmonics to the particlization model. One of the main goals of heavy-ion physics is to extract quantitative values for properties of the quark-gluon plasma (QGP), such as its specific shear viscosity $\eta/s$ by matching collision simulations to experimental data. To estimate the sensitivity, we multiply the local shear stress tensor at each point on the conversion hypersurface by a constant factor $\kappa$ in order to mimic a change in specific shear viscosity $\eta / s$ $\to$ $\kappa \eta / s$. In the Navier-Stokes approximation (\ref{pi_est}), this is reasonable because viscous corrections to flow gradients are generally small \cite{teaneyv2,IStrv}. We then capture the difference in particlization models \NEW{empirically} via adjusting $\kappa$.

\begin{figure}[h!]
\includegraphics[width=0.49\linewidth]{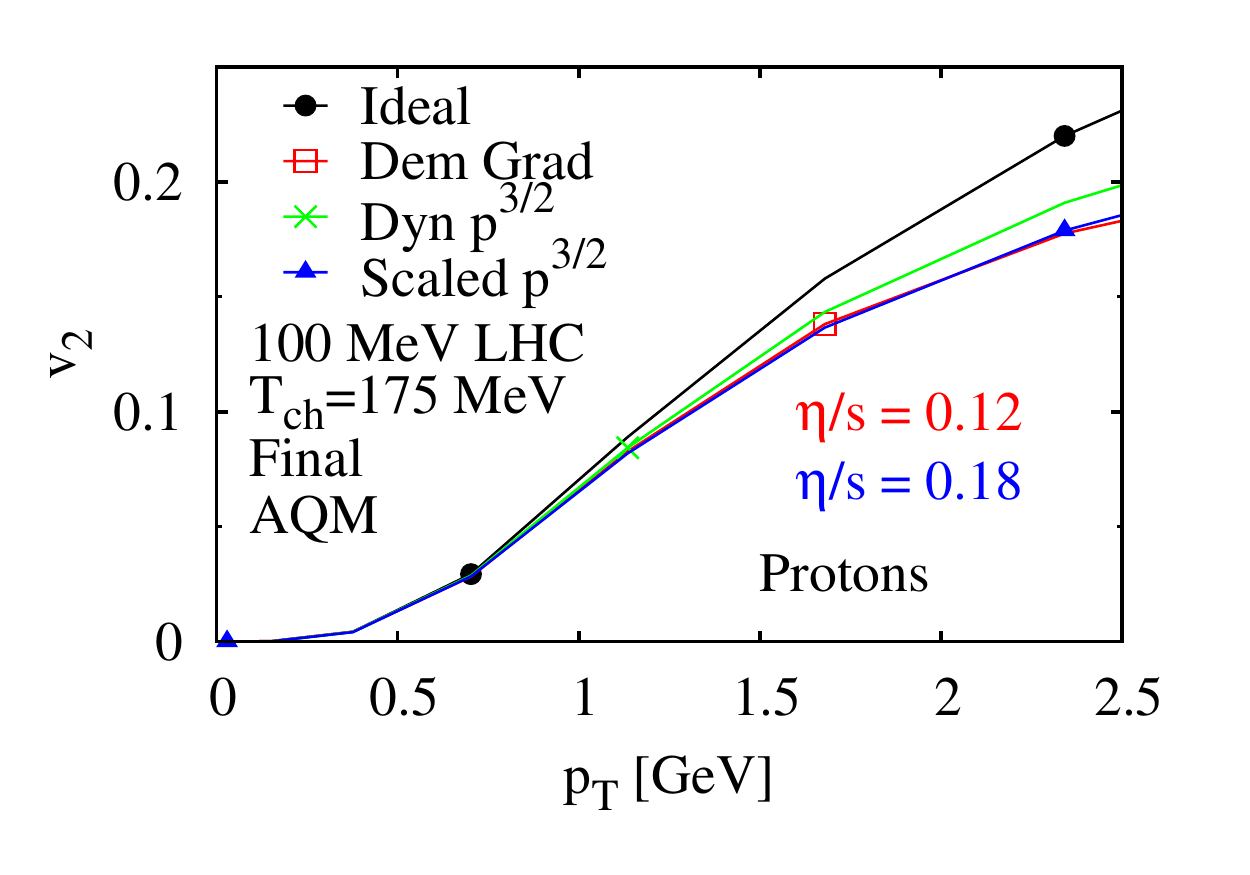}
\caption{Differential elliptic flow $v_2(p_T)$ of protons in $Pb + Pb$ at $\sqrt{s_{NN}}=2.76$~TeV at the LHC at 20-30\% centrality using 2+1D boost invariant viscous hydrodynamic solutions \cite{PaatelainenNiemi:2014} and fluid-to-particle conversion at $T_{conv} = 100$~MeV. Feed-down from decays of unstable resonances has been included. The standard democratic Grad approach (open boxes) is compared to self-consistent shear corrections with momentum dependence $\delta f_i \propto p^{3/2}$ computed for a 49-species effective hadron gas from linearized kinetic theory with unscaled shear stress (crosses) and shear stress multiplied by $\kappa =1.5$ everywhere on the conversion hypersurface (filled triangles). Results with uncorrected, local equilibrium phase-space distributions ($\delta f = 0$) are also shown (filled circles) for reference.}
\label{Fig:LHC-v2-extract}
\end{figure}

Figure \ref{Fig:LHC-v2-extract} demonstrates that both self-consistent particlization and the democratic Grad ansatz can reproduce the same proton $v_2(p_T)$ curve from the calculation in the previous Section for $Pb + Pb$ collisions at the LHC, provided one increases $\eta / s$ by 50\% (i.e., $\kappa = 1.5$) in the self-consistent case. Three of the proton $v_2(p_T)$ curves shown are the same as in Fig. \ref{Fig:LHC-v2-3o2}, computed with ideal freezeout (filled circles), democratic Grad (open boxes), and the self-consistent approach (crosses). The fourth curve (filled triangles) is the self-consistent result, but with shear stress scaled by $\kappa = 1.5$, making it practically identical to the democratic Grad curve. In the original hydrodynamic simulation the specific shear viscosity was $\eta / s = 0.12$ in the plasma phase, which comprises a good portion of the hydrodynamic evolution. If, with standard democratic freezeout, the calculation fits experimental data then one might infer an effective $\eta / s = 0.12$ for the QGP. However, with self-consistent freezeout a 50\% larger $\eta / s \approx 0.18$ would be needed to agree with the data.

%%%
\subsection{Shear viscosity comparison}
\label{Sc:compare}

\begin{figure}[h!]
\includegraphics[width=0.49\linewidth]{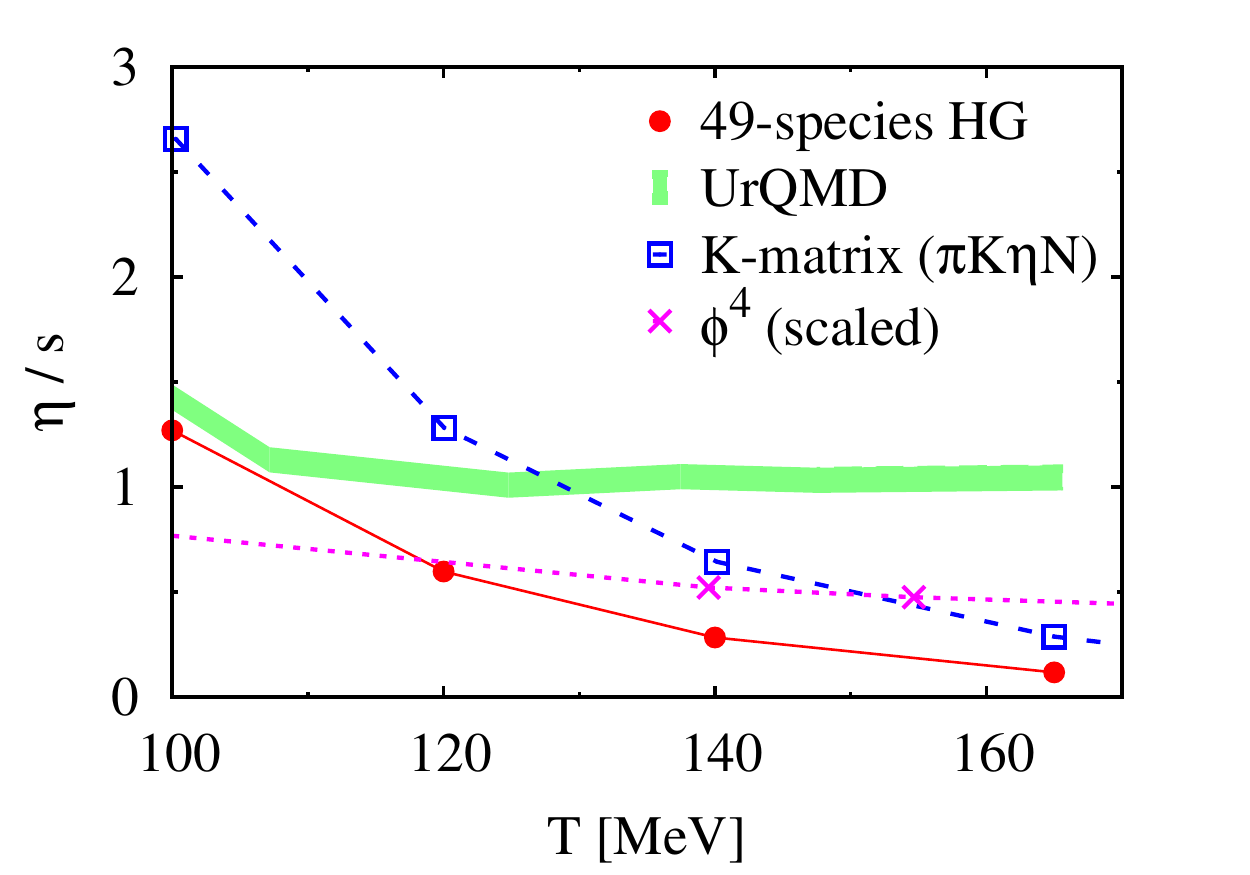}
\caption{Comparison of the specific shear viscosity $\eta/s$ as a function of temperature
from a variety of models, for mixtures at zero baryon density:
i) the self-consistent kinetic theory calculation employed in this work 
for a hadron gas of 49 effective 
species (solid red curve, filled circles), ii) the hadron transport model 
UrQMD\cite{UrQMD}, extracted in Ref. \cite{Demir:2008tr} (shaded green band), 
and iii) for a $\pi-K-\eta-N$
mixture, from the $K$-matrix approach in Ref. \cite{Wiranata:2013oaa} (dashed blue line,
with squares).
For illustration (dotted magenta line, with crosses), we also plot the approximate 
temperature dependence of $\eta/s$ in $\lambda \phi^4$ theory at weak coupling for $T \gg m$, based on
\cite{Jeon:1995zm} (see text).}
\label{Fig:etas-compare}
\end{figure}

\NEW{It is interesting to compare our hadron gas shear viscosity calculation
to other works in the literature, in particular,
results by
Demir {\it et al} \cite{Demir:2008tr} extracted from the hadron transport code
UrQMD \cite{UrQMD},
and calculations by Wiranata {\it et al} based on
the $K$-matrix approach \cite{Wiranata:2013oaa}.
Figure \ref{Fig:etas-compare} shows the specific shear viscosities $\eta/s$
in the temperature window $100~{\rm MeV} < T < 165~{\rm MeV}$ from these
approaches, for hadronic mixtures at zero baryon density ($\mu_B = 0$).
Linearized kinetic theory used in this work 
gives for a hadron gas of 49 effective species
an $\eta/s$ that drops markedly with
temperature,
from about $\eta/s \approx 1.3$ at $T=100$~MeV to $0.2$ at $165$~MeV
(solid red curve, with filled circles).
The monotonic decrease is driven almost entirely by the denominator of
$\eta/s$, i.e., the monotonic increase of the
hadron gas entropy density with temperature
\be
s = \sum_i s_i = \sum_i \frac{e_i + P_i}{T}
  = \sum_i \frac{g_i}{2\pi^2} m_i^3 T K_3\!\left(\frac{m_i}{T}\right)
\label{HG_entropy}
\ee
where $g_i$ is the degeneracy factor, $K_n$ is a modified Bessel function of the second kind, and Eq. (\ref{HG_entropy}) is valid in the Boltzmann limit.
The hadron gas shear viscosity
actually increases with temperature in our approach,
albeit {\em rather} slowly.
Moreover, we find that hadron gas shear viscosity is dominated by the lightest species, 
most importantly the pions, and changes only little as more and more species are included
in the calculation.}

\NEW{A qualitatively similar drop in $\eta/s$ with temperature can be seen 
in Fig.~\ref{Fig:etas-compare} 
from $K$-matrix cross sections \cite{Wiranata:2013oaa}, for a mixture of pions, kaons,
$\eta(548)$-s and nucleons (dashed blue line, with squares).
Quantitatively, however, the $K$-matrix result is twice as large as
our calculation. This is not surprising. We both agree on the hadron gas
entropy density; in particular, the entropy density we extract from
their Figs. 13 and 14, via $s = \eta / (\eta/s)$, matches our result for a 
$\pi-K-\eta-N$ mixture
to within a couple percent. Where the two calculations disagree is 
the hadron gas viscosity. We used effective hadronic cross
sections chosen to reproduce mean scattering times for pions, kaons, and nucleons
calculated by Prakash {\it et al} in \cite{Prakash:1993bt}. Compared to
the shear viscosity from the $K$-matrix approach, the shear viscosity
calculated in Ref. [3] is smaller by nearly a factor of
2 (compare Figs. 10 and 11 in Ref. \cite{Prakash:1993bt} 
to Fig. 13 in \cite{Wiranata:2013oaa}). 
We do reproduce the shear viscosity in Ref. \cite{Prakash:1993bt} 
to better than $30$\% (not shown) 
in the temperature range $100$~MeV $ < T < 165$~MeV studied here.}

\NEW{In light of the two kinetic theory calculations discussed above 
it is rather striking that UrQMD 
gives \cite{Demir:2008tr}
an essentially flat hadron gas $\eta/s$ versus temperature 
(shaded  green band in Fig.~\ref{Fig:etas-compare}).
It would be useful to investigate in more detail in the future 
whether $\eta/s$ from UrQMD comes out largely
independent of temperature because entropy density in UrQMD increases much slower than 
for an ideal gas of hadrons, or whether it is the
shear viscosity that increases in UrQMD much more rapidly with temperature. 
It should be noted that the dynamics of UrQMD includes not only particles (hadrons) 
but also extended objects (strings), which might be responsible for this unusual behavior.}

\NEW{Finally, it is illustrative to provide a rough comparison to the temperature dependence 
of $\eta/s$ from
$\lambda \phi^4$ theory at weak coupling
(dotted magenda line, with crosses). Here we use the 
shear viscosity calculation by Jeon {\it et al} \cite{Jeon:1995zm} that gives
the viscosity in units of the thermal mass, i.e., $\eta/m_{th}^3$,
versus normalized temperature $T/m_{th}$ (cf. Fig. 4 therein),
and we ignore
interaction corrections both in the thermal mass $m_{th}^2 = m^2 + {\cal O}(\lambda T^2)$
and the entropy (i.e., we use Eq.~(\ref{HG_entropy})). 
The scalar mass $m = 0.14$~GeV is set to the pion mass. 
At weak coupling, shear viscosity in $\lambda\phi^4$
theory is very large; for $T \gg m$,
$\eta \sim 3000 T^3/\lambda^2 \ggg s$. Therefore, we divide $\eta/s$ 
by an arbitrary constant factor to highlight its temperature dependence. 
The end result is a modest,
monotonic {\em decrease} in $\eta/s$ by about one-third from $T = 100$~MeV
to $T=165$~MeV,
in qualitative agreement with the dropping trend seen earlier in kinetic 
kinetic theory. The decrease is weaker partly because in scalar theory the $2\to 2$ 
cross section $\sigma \sim \lambda^2/32\pi s$ drops with energy, which makes the
shear viscosity increase more rapidly with temperature.}

%%%
\subsection{Higher flow harmonics}
\label{Sc:v4v6}

\begin{figure}[h!]
\includegraphics[width=0.49\linewidth]{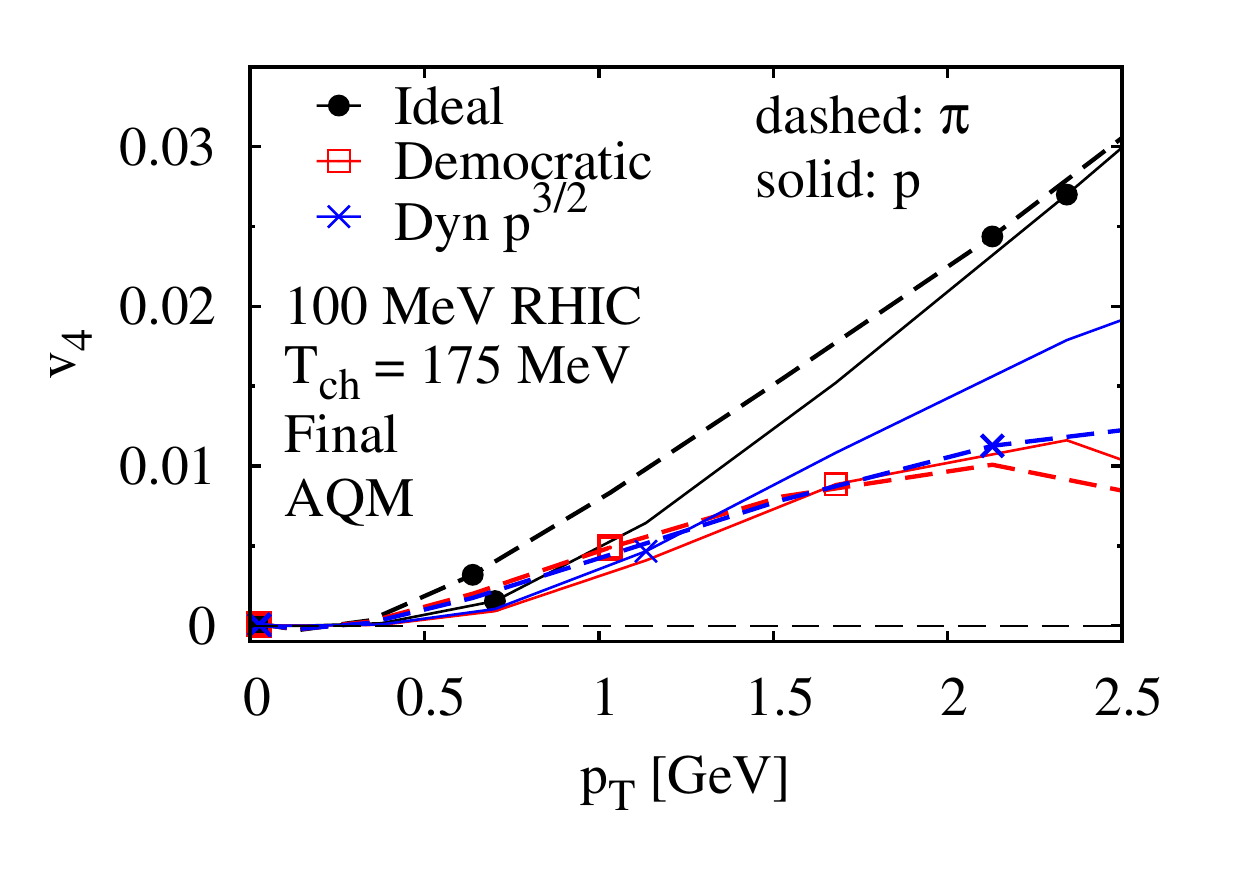}
\caption{Differential $4^{th}$ flow harmonic $v_4(p_T)$ of pions and protons at RHIC from the same calculation as in Fig. \ref{Fig:RHIC-v2-3o2}.}
\label{Fig:RHIC-v4-3o2}
\end{figure}

In systems with nonzero shear viscosity, velocity gradients in general smooth out between adjacent layers of the fluid. Higher flow coefficients with $n > 2$ in Eq. (\ref{vn}) encode anisotropies at progressively smaller angular separations, and thus tend to get evened out more efficiently than elliptic flow \cite{Alver:2010}. Figure \ref{Fig:RHIC-v4-3o2} shows the differential $4^{th}$ flow harmonic $v_4(p_T)$ for pions and protons in $Au+Au$ at RHIC for the same calculation shown in Fig. \ref{Fig:RHIC-v2-3o2}. 
Qualitative features of $v_4(p_T)$ are similar to those of $v_2(p_T)$, such as mass ordering and crossing of pion and proton flow, and the viscous suppression relative to results from ideal freezeout. The sensitivity to the particlization model used is, however, stronger than in $v_2$. For viscous freezeout the crossing between protons and pions occurs at a noticeably lower $p_T$ for $v_4$ than for $v_2$. This narrows the $p_T$ range for traditional mass splitting with $v_4^p < v_4^\pi$. Also, viscous freezeout with the standard democratic Grad ansatz (open boxes) reduces $v_4$ by at least a factor of two at $p_T > 1.8$~GeV compared to ideal freezeout (filled circles). On the other hand, the self-consistent approach (crosses) suppresses proton $v_4$ less, leading to a large pion-proton difference of nearly 50\% at $p_T \approx 2.5$~GeV. 

\begin{figure}[h!]
\includegraphics[width=0.49\linewidth]{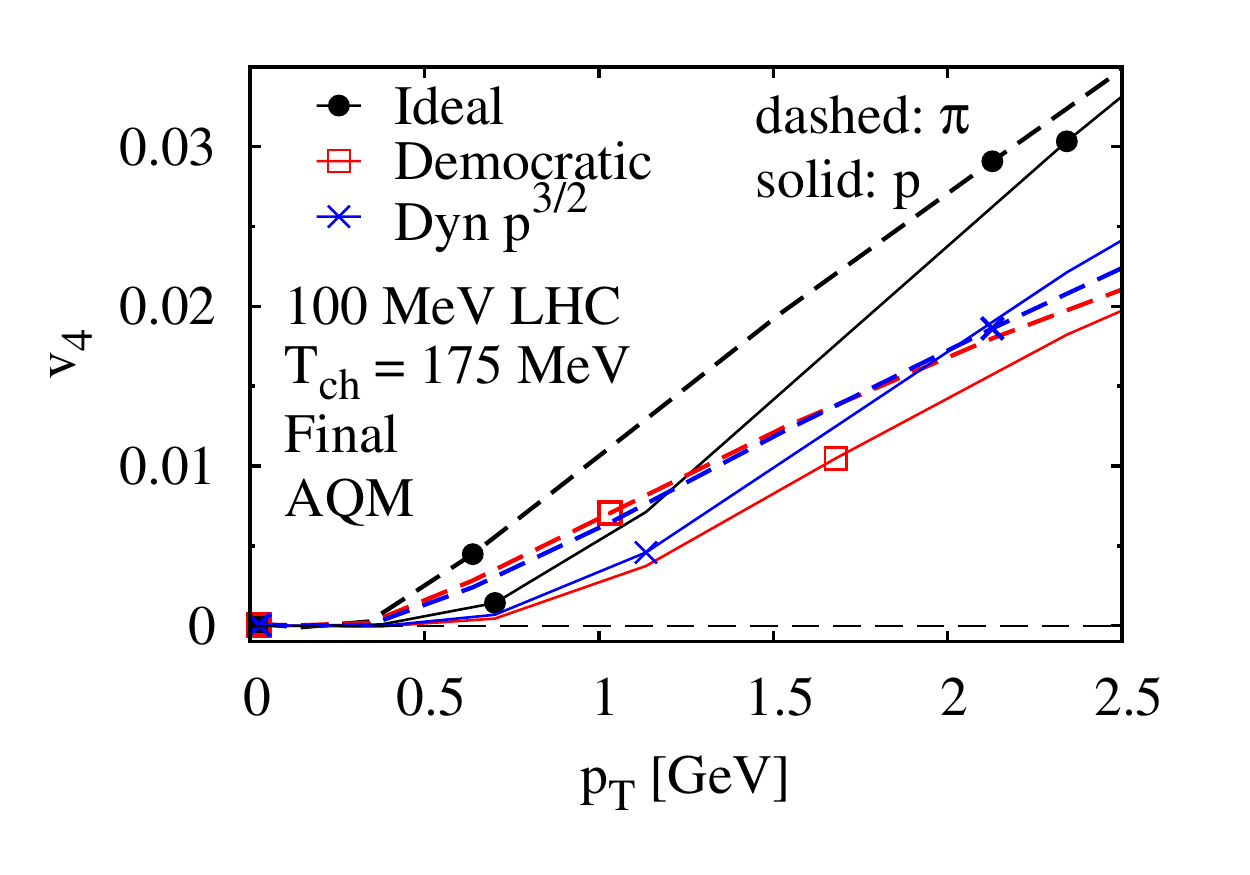}
\includegraphics[width=0.49\linewidth]{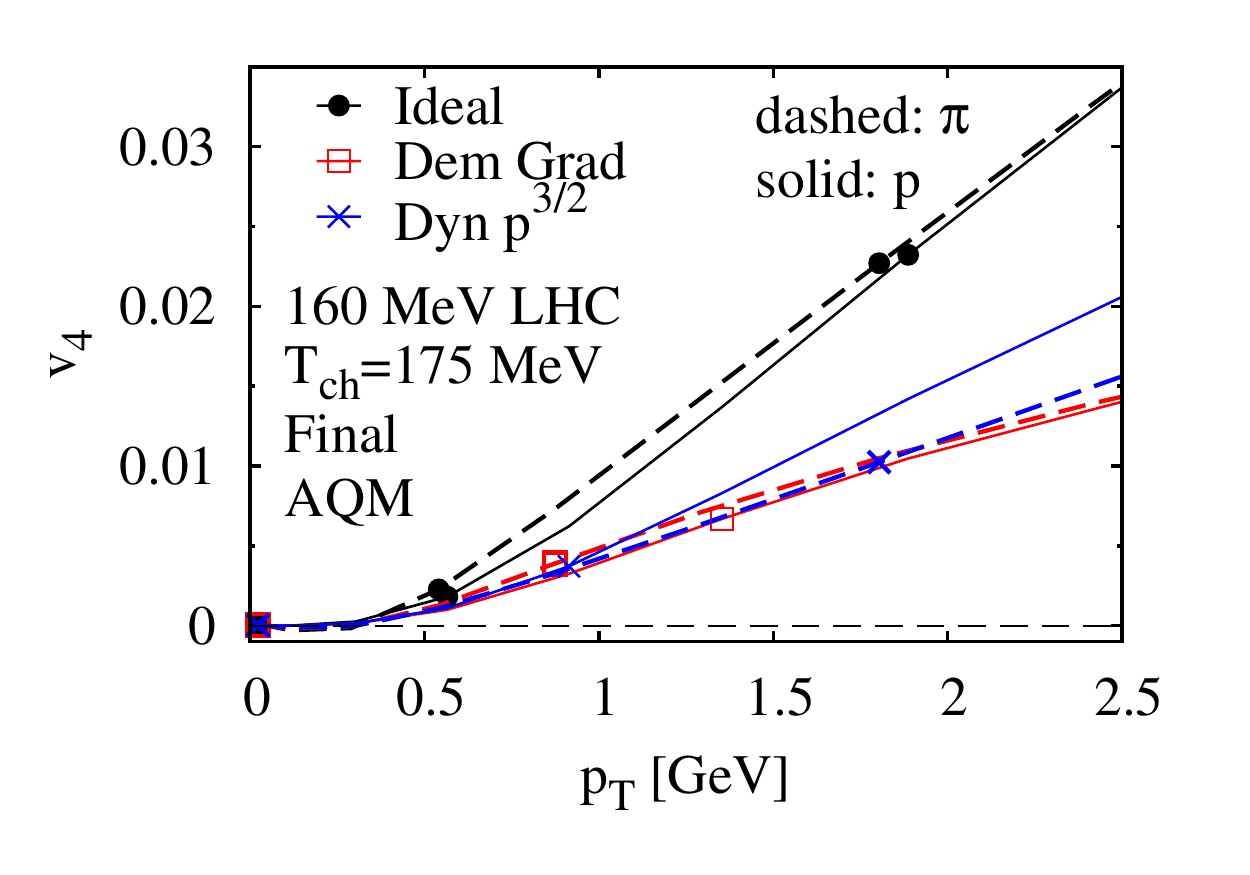}
\caption{Differential $4^{th}$ flow harmonic $v_4(p_T)$ of pions and protons at the LHC from the same calculation as in Fig. \ref{Fig:LHC-v2-3o2}.}
\label{Fig:LHC-v4-3o2}
\end{figure}

Figure \ref{Fig:LHC-v4-3o2} shows $v_4(p_T)$ for pions and protons in $Pb + Pb$ at the LHC for the same calculations shown in Fig. \ref{Fig:LHC-v2-3o2}. For $T_{conv} = 100$~MeV (left panel), which is relevant for direct comparison between hydrodynamics and LHC data, there is a larger separation at low $p_T$ between the flows of the two species and the various particlization models than in the corresponding $v_4(p_T)$ results at RHIC. At high $p_T$, however, the spread in $v_4$ is smaller at the LHC. Remarkably, for the higher conversion temperature $T_{conv} = 160$~MeV relevant for hybrid (hydro+transport) calculations, all $v_4(p_T)$ curves at the LHC look similar, even quantitatively, to those in $Au+Au$ at RHIC with $T_{conv} = 100$~MeV. This, in part, must be a reflection of the shorter evolution time, during which smaller hydrodynamic flow is generated. One should also note that using hypersurfaces from realistic viscous hydrodynamic simulations has a dramatic effect on $v_4(p_T)$, as this observable was roughly zero for protons and negative for pions when estimated from ideal hydrodynamic evolution in Ref. \cite{Wolff:2014}.

\begin{figure}[h!]
\includegraphics[width=0.49\linewidth]{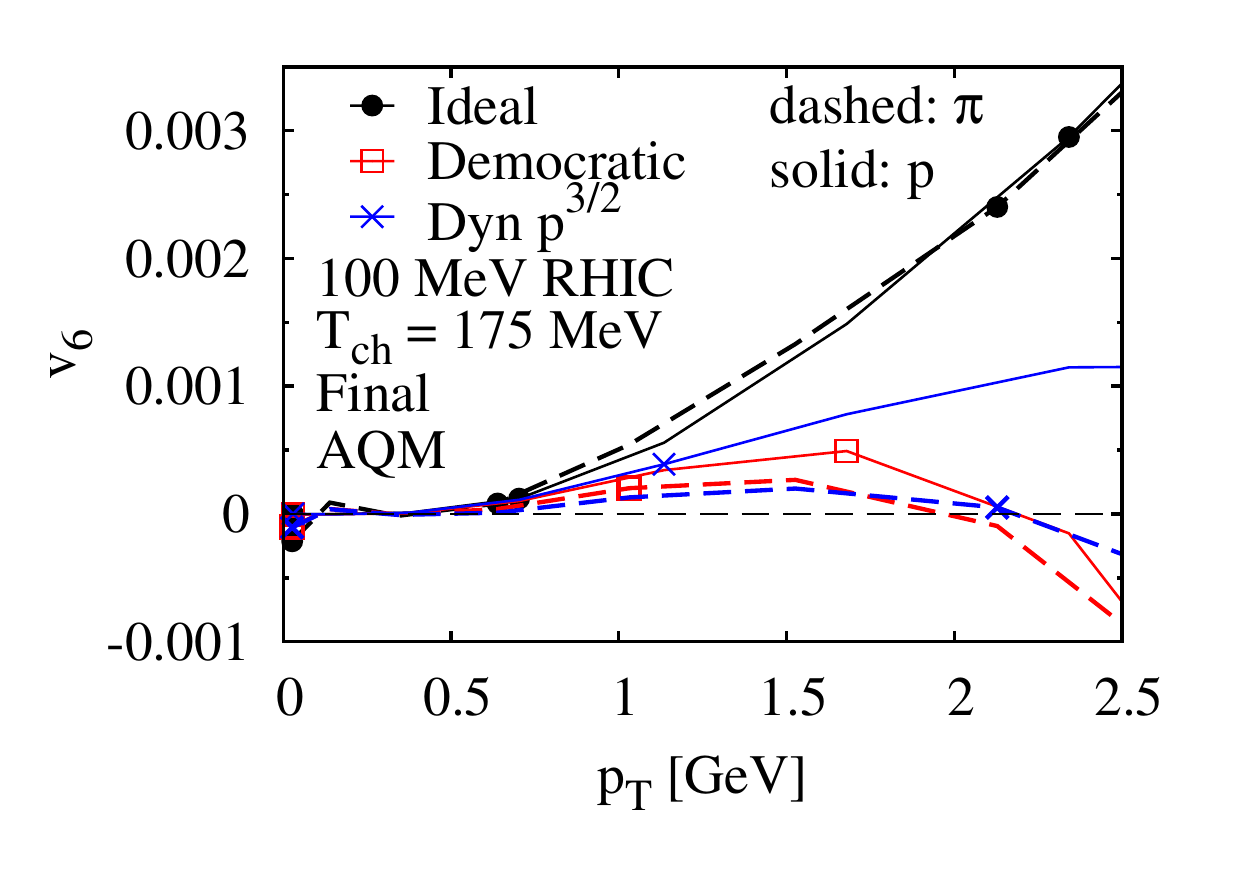}
\caption{Differential $6^{th}$ flow harmonic $v_6(p_T)$ of pions and protons at RHIC from the same calculation as in Figs. \ref{Fig:RHIC-v2-3o2} and \ref{Fig:RHIC-v4-3o2}.}
\label{Fig:RHIC-v6-3o2}
\end{figure}

\begin{figure}[h!]
\includegraphics[width=0.49\linewidth]{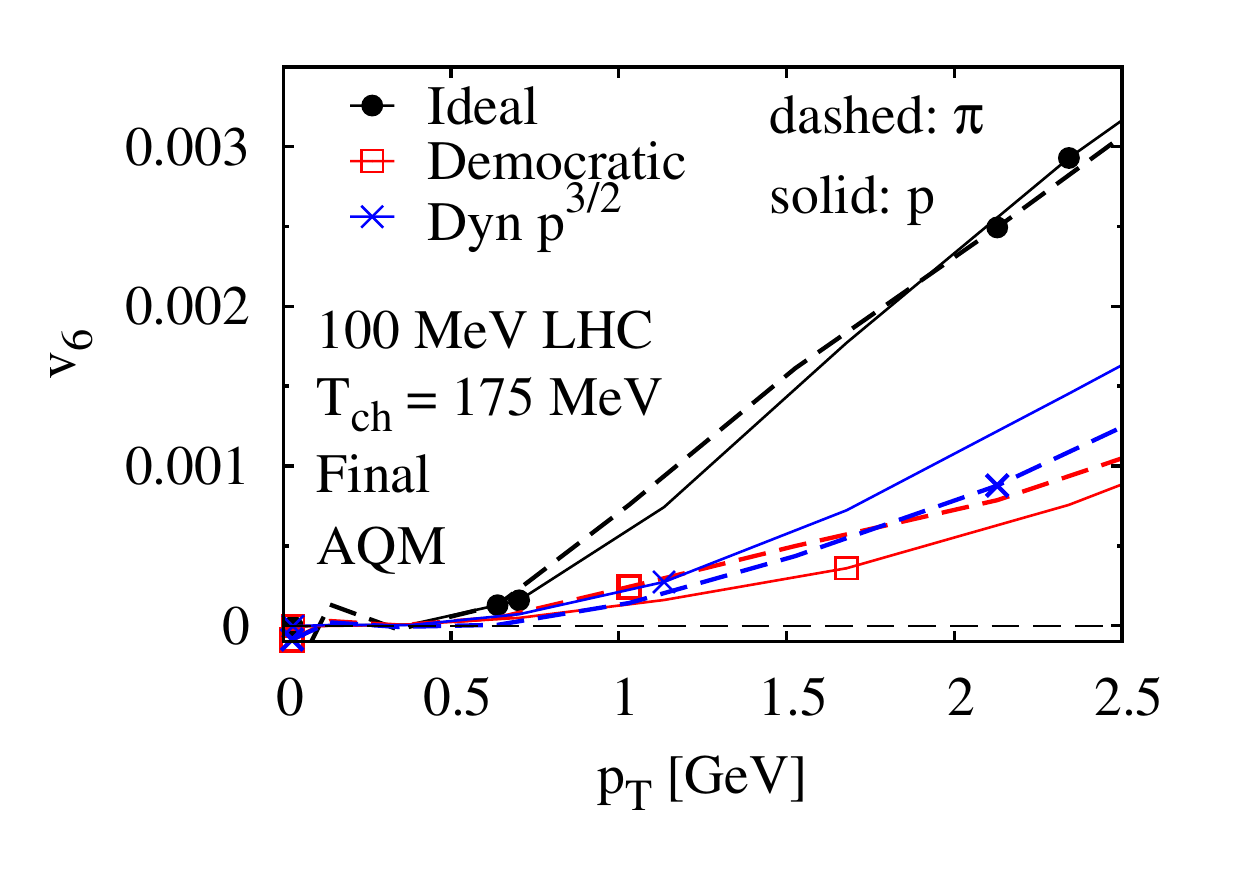}
\includegraphics[width=0.49\linewidth]{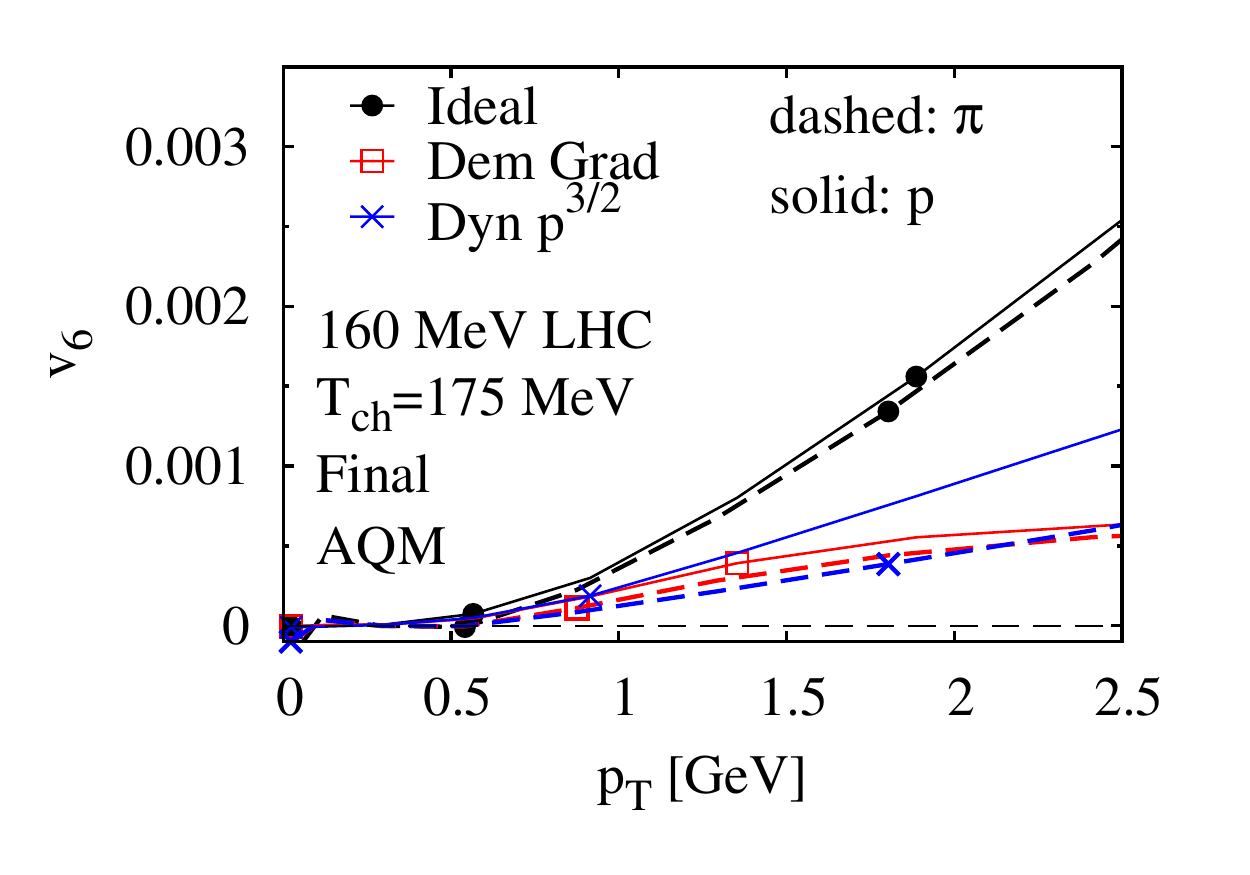}
\caption{Differential $6^{th}$ flow harmonic $v_6(p_T)$ of pions and protons at the LHC from the same calculation as in Figs. \ref{Fig:LHC-v2-3o2} and \ref{Fig:LHC-v4-3o2}.}
\label{Fig:LHC-v6-3o2}
\end{figure}

Finally, Figs. \ref{Fig:RHIC-v6-3o2} and \ref{Fig:LHC-v6-3o2} show $v_6(p_T)$ for pions and protons in $Au+Au$ at RHIC and $Pb+Pb$ at the LHC, respectively. As expected, viscous corrections to $v_6$ are generally larger than for $v_4$, exceeding even a factor of three in some cases. In $Au+Au$ collisions at RHIC, viscosity reduces $v_6$ to nearly zero; in fact, $v_6$ goes negative for $p_T > 2$~GeV, except for protons from self-consistent freezeout which maintain a positive $v_6$ in the entire $p_T$ range shown. In contrast, viscous corrections for $Pb+Pb$ collisions at the LHC, though large, leave $v_6$ positive in all cases studied here. For the higher conversion temperature of 160 MeV (Fig. \ref{Fig:LHC-v6-3o2} right panel), pion and proton $v_6$ from democratic Grad freezeout are largely identical. While with self-consistent conversion to particles, pion $v_6$ stays about the same but proton $v_6$ nearly doubles, resulting in a two-to-one proton to pion $v_6$ ratio. The self-consistent and democratic Grad particlizations both give much the same $v_6$ for pions at $T_{conv} = 100$~MeV (Fig. \ref{Fig:LHC-v6-3o2} left panel) as well. However, they differ in proton $v_6$; specifically, proton $v_6$ is $\approx 20\%$ higher than pion $v_6$ from the self-consistent approach, whereas it is $\approx 15-30\%$ below pion $v_6$ with democratic Grad corrections.

The results of this section cleary underscore the need for careful future comparisons between precise identified particle flow harmonics data from both RHIC and the LHC up to $p_T = 2-3$~GeV and state of the art hydrodynamic and hybrid calculations that employ realistic fluctuating initial conditions together with self-consistent particlization.

%%%%%%%%%%
\section{Conclusions}
\label{Sc:Conc}

The quantitative extraction of QGP properties such as the specific shear viscosity in the viscous hydrodynamic paradigm inevitably requires the conversion of a dissipative fluid to particles. This so-called particlization is typically done using the Cooper-Frye formula (\ref{CooperFrye}) with hadron phase-space densities $f_i = f_i^{eq} + \delta f_i$ that include corrections to thermal distributions which are chosen to be independent of particle dynamics and quadratic in momentum (democratic Grad ansatz). This naive approximation completely ignores the dynamics that keeps the hadron gas near equilibrium. Here, self-consistent shear viscous corrections are calculated from linearized kinetic theory using hadronic cross sections motivated by the additive quark model \cite{AQM}. 
The corrections were then used to compute differential harmonic flow coefficients $v_2(p_T)$, $v_4(p_T)$, and $v_6(p_T)$ in $Au+Au$ collisions at top RHIC energy $\sqrt{s_{NN}} = 200$~GeV and $Pb+Pb$ collisions at the LHC energy $\sqrt{s_{NN}} = 2.76$~TeV. Expanding upon previous works \cite{MolnarWolff,Wolff:2014}, we include early chemical freezeout in the hadron gas and use Cooper-Frye hypersurfaces from real viscous hydrodynamic evolution. 

We find that self-consistent particlization leads to larger proton elliptic flow at moderately high $p_T \sim 2-3$ GeV compared to that of pions, qualitatively corroborating our prior estimate \cite{MolnarWolff} of the effect on elliptic flow. In addition, we show that $v_4$ and $v_6$ are more sensitive than $v_2$ to the hadron distributions used in the conversion. In fact, with self-consistent, species-dependent viscous corrections, the pion-proton splitting in $v_4(p_T)$ and $v_6(p_T)$ can be surprising large. For example, in $Au+Au$ at RHIC, pion and proton $v_6$ can even have different signs at moderately high transverse momentum. 

The ambiguity in particlization model leads to a theoretical uncertainty in the  specific shear viscosity of the quark-gluon plasma extraced from elliptic flow data. For $Pb+Pb$ collisions at the LHC, we estimate the uncertainty to be as high as 50\%.

We note that there are several simplifications made in this work. The use of constant cross sections instead of the realistic energy-dependent cross sections between hadron species will need to be remedied in a future study. In addition, the momentum dependence of relative viscous corrections, $\phi_i \equiv \delta f_i / f^{eq}_i$, was approximated here by a single $p^{3/2}$ power. 
Nevertheless, our results indicate the need for careful comparisons between hydrodynamic calculations and precise data on identified-particle $v_4(p_T)$ and $v_6(p_T)$ up to $p_T \sim 2-3$~GeV. To facilitate such studies, we give in Table \ref{Table:Cfits-Tch175} convenient parametrizations of the self-consistent shear viscous corrections for each hadron species.

%%%
\acknowledgments 
The authors thank Harri Niemi for the viscous hydrodynamic hypersurfaces used in this work. Insightful discussions with Gabriel Denicol, Chun Shen, Derek Teaney, Guy Moore, Sangyong Jeon, and Raju Venugopalan are also acknowledged. Z.W. thanks the Institut f\"ur Theoretische Physik at Goethe University (Frankfurt, Germany) for their generosity where parts of this work were done. D.M. thanks RIKEN, Brookhaven National Laboratory and the US Department of Energy for providing facilities essential for the completion of this work. D.M. also thanks the hospitality of the Wigner Research Center for Physics (Budapest, Hungary) where parts of this work have been done. Computing resources managed by RCAC/Purdue are also gratefully acknowledged. This work was supported by the U.S. Department of Energy, Office of Science, under grants DE-AC02-98CH10886 [RIKEN BNL] and DE-SC0004035.

%%%%%%%%%

\end{document}